\documentclass[11pt]{revtex4}     
\usepackage{amssymb,epsf}
\usepackage{latexsym}
\begin{document}
\title{Thermodynamics of viscous dark energy in RSII braneworld}
\author{M. R. Setare $^{1,2}$\footnote{rezakord@ipm.ir} and  A. Sheykhi$^{2,3}$\footnote{sheykhi@mail.uk.ac.ir}}
\address{$^1$Department of Science, Payame Noor University, Bijar, Iran
\\
         $^2$Research Institute for Astronomy and Astrophysics of Maragha (RIAAM), Maragha,
         Iran\\
$^{3}$ Department of Physics, Shahid Bahonar University, P.O. Box
76175, Kerman, Iran }

\begin{abstract}
We show that for a RSII braneworld filled with interacting viscous
dark energy and dark matter, one can always rewrite the Friedmann
equation in the form of the first law of thermodynamics,
$dE=T_hdS_h+WdV$, at apparent horizon. In addition, the generalized
second law of thermodynamics can fulfilled in a region enclosed by
the apparent horizon on the brane for both constant and time
variable 5-dynamical Newton's constant $G_5$. These results hold
regardless of the specific form of the dark energy. Our study
further support that in an accelerating universe with spatial
curvature, the apparent horizon is a physical boundary from the
thermodynamical point of view.
\end{abstract}
\maketitle

\section{Introduction}\label{Intr}
Observational data indicates that our universe is currently under
accelerating expansion \cite{{1},{111}}. It seems that some unknown
energy components (dark energy) with negative pressure are
responsible for this late-time acceleration \cite{2}. However,
understanding the nature of dark energy is one of the fundamental
problems of modern theoretical cosmology \cite{3}. An alternative
approach to accommodate dark energy is modifying the general theory
of relativity on large scales. Among these theories, scalar-tensor
theories \cite{4}, f(R) gravity \cite{5}, DGP braneworld gravity
\cite{6} and string-inspired theories
\cite{7} are studied extensively.\\
The cosmological models with non-viscous cosmic fluid has been
studied widely in the literature. Early treatises on viscous
cosmology are given in \cite{Pad}. The viscous entropy production in
the early universe and viscous fluids on the Randall-Sundrum branes
have been studied respectively in \cite{Bre0}. A special branch of
viscous cosmology is to investigate how the bulk viscosity can
influence the future singularity, commonly called the Big Rip, when
the fluid is in the phantom state corresponding to $w_D<-1$. A lot
of works have been done in this direction \cite{Bre1,Bre2}. In
particular, it was first pointed out in \cite{Bre1} that the
presence of a bulk viscosity proportional to the Hubble expansion
$H$ can cause the fluid to pass from the quintessence region into
the phantom region and thereby inevitably lead to a future
singularity.\\
 In the present work we are interested to investigate
the
 interacting viscous dark energy and dark matter in RSII braneworld, from the
thermodynamic point of view. In particular, we desire to examine
under what conditions the underlying system obeys the generalized
second law of thermodynamics, namely the sum of entropies of the
individual components, including that of the background, to be
positive. Then we extend our analysis with considering the time
variable $5$D Newton's constant $G_5$.  Until now, in most the
investigated dark energy models a constant Newton's ``constant'' $G$
has been considered. However, there are significant indications that
$G$ can by varying, being a function of time or equivalently of the
scale factor \cite{G4com}. In particular, observations of
Hulse-Taylor binary pulsar \cite{Damour,kogan}, helio-seismological
data \cite{guenther}, Type Ia supernova observations \cite{1}  and
astereoseismological data from the pulsating white dwarf star
G117-B15A \cite{Biesiada} lead to $\left|\dot{G}/G\right|
\lessapprox 4.10 \times 10^{-11} yr^{-1}$, for $z\lesssim3.5$
\cite{ray1}. Additionally, a varying $G$ has some theoretical
advantages too, alleviating the dark matter problem \cite{gol}, the
cosmic coincidence problem \cite{jamil} and the discrepancies in
Hubble parameter value \cite{ber}.

There have been many proposals in the literature attempting to
theoretically justified a varying gravitational constant, despite
the lack of a full, underlying quantum gravity theory. Starting with
the simple but pioneering work of Dirac \cite{Dirac:1938mt}, the
varying behavior in Kaluza-Klein theory was associated with a scalar
field appearing in the metric component corresponding to the  $5$-th
dimension \cite{kal} and its size variation \cite{akk}. An
alternative approach arises from Brans-Dicke framework \cite{bd},
where the gravitational constant is replaced by a scalar field
coupling to gravity through a new parameter, and it has been
generalized to various forms of scalar-tensor theories \cite{gen},
leading to a considerably broader range of variable-$G$ theories. In
addition, justification of a varying Newton's constant has been
established with the use of conformal invariance and its induced
local transformations \cite{bek}. Finally, a varying $G$ can arise
perturbatively through a semiclassical treatment of Hilbert-Einstein
action \cite{19}, non-perturbatively through quantum-gravitational
approaches within the ``Hilbert-Einstein truncation'' \cite{21}, or
through gravitational holography \cite{Guberina,71}.
\section{Basic Equations}\label{RS}
Our starting point is the four-dimensional homogenous and
isotropic FRW universe on the brane with the metric
\begin{equation}
ds^2={h}_{\mu \nu}dx^{\mu}
dx^{\nu}+\tilde{r}^2(d\theta^2+\sin^2\theta d\phi^2),
\end{equation}
where $\tilde{r}=a(t)r$, $x^0=t, x^1=r$, the two dimensional metric
$h_{\mu \nu}$=diag $(-1, a^2/(1-kr^2))$. Here $k$ denotes the
curvature of space with $k = 0, 1, -1$ corresponding to open, flat,
and closed universes, respectively. A closed universe with a small
positive curvature ($\Omega_k\simeq0.01$) is compatible with
observations \cite{wmap}. The dynamical apparent horizon, a
marginally trapped surface with vanishing expansion, is determined
by the relation $h^{\mu \nu}\partial_{\mu}\tilde
{r}\partial_{\nu}\tilde {r}=0$, which implies that the vector
$\nabla \tilde {r}$ is null on the apparent horizon surface. The
apparent horizon was argued as a causal horizon for a dynamical
spacetime and is associated with gravitational entropy and surface
gravity \cite{Hay2,Bak}. For the FRW universe the apparent horizon
radius reads
\begin{equation}
\label{radius}
 \tilde{r}_A=\frac{1}{\sqrt{H^2+k/a^2}}.
\end{equation}
The associated surface gravity on the apparent horizon can be
defined as
\begin{equation}
\label{surgra}\label{kappa}
 \kappa =\frac{1}{\sqrt{-h}}\partial_{a}\left(\sqrt{-h}h^{ab}\partial_{ab}\tilde
 {r}\right),
\end{equation}
thus one can easily express the surface gravity on the apparent
horizon
\begin{equation}\label{surgrav}
\kappa=-\frac{1}{\tilde r_A}\left(1-\frac{\dot {\tilde
r}_A}{2H\tilde r_A}\right).
\end{equation}
The associated temperature on the apparent horizon can be
expressed in the form
\begin{equation}\label{Therm}
T_{h} =\frac{|\kappa|}{2\pi}=\frac{1}{2\pi \tilde
r_A}\left(1-\frac{\dot {\tilde r}_A}{2H\tilde r_A}\right).
\end{equation}
where $\frac{\dot{\tilde r}_A}{2H\tilde r_A}<1$ ensures that the
temperature is positive. Recently the Hawking radiation on the
apparent horizon has been observed in \cite{cai3} which gives more
solid physical implication of the temperature associated with the
apparent horizon.

The Friedmann equation for $3$-dimensional Randall-Sundrum (RS) II
brane embedded in an $5$-dimensional AdS bulk can be written
\cite{Bin}
\begin{equation}\label{RSFri}
H^2+\frac{k}{a^2}-\frac{\kappa_{5}^2\Lambda_{5}}{6}-\frac{\mathcal{C}}{a^4}
=\frac{\kappa_{5}^4}{36}\rho^2.
\end{equation}
where \begin{equation}\label{rela}
 \kappa_{5}^2=8\pi
 G_{5}\, ,\quad
 \Lambda_{5}=-\frac{6}{\kappa_{5}^2\ell^2},
\end{equation}
$\Lambda_{5}$ is the $5$-dimensional bulk cosmological constant, and
$\ell$ is the AdS radius of the bulk spacetime. Here
$\rho=\rho_m+\rho_D$ where $\rho_m$ and $\rho_D$ are, respectively,
the energy density of dark matter and dark energy confined to the
brane and $H=\dot{a}/a$ is the Hubble parameter on the brane. The
constant $\mathcal{C}$ comes from the $5$-dimensional bulk Weyl
tensor. In this paper we are interested in AdS bulk spacetimes, so
the bulk Weyl tensor vanishes and thus we set $\mathcal{C}=0$ in the
following discussions. The energy-momentum tensor of  the matter and
energy content on the brane is as,
\begin{equation}\label{T}
T_{\mu\nu}=\rho
u_{\mu}u_{\nu}+\tilde{p}_D(g_{\mu\nu}+u_{\mu}u_{\nu}),
\end{equation}
where $u_{\mu}$ is the four-velocity vector, and
\begin{equation}\label{cons}
\tilde{p}_D={p}_D-3H\xi,
\end{equation}
is the effective pressure of dark energy and $\xi$ is the viscosity
coefficient. The condition $\xi>0$ guaranties a positive entropy
production and, in consequence, no violation of the second law of
the thermodynamics \cite{Zim}. The total energy density on the brane
satisfies a conservation law
\begin{equation}\label{cons}
\dot{\rho}+3H(\rho+\tilde{p}_D)=0.
\end{equation}
However, since we consider the interaction between dark matter and
dark energy, $\rho_{m}$ and $\rho_{D}$ do not conserve separately,
they must rather enter the energy balances
\begin{eqnarray}
&&\dot{\rho}_m+3H\rho_m=Q, \label{consm}
\\&& \dot{\rho}_D+3H\rho_D(1+w_D)=9H^2\xi-Q.\label{consq}
\end{eqnarray}
where $w_{D}=p_D/\rho_D$ is the equation of state parameter of
viscous dark energy and $Q=\Gamma \rho_D$ denotes the interaction
between the dark components. We also assume the interaction term is
positive, $Q>0$, which means that there is an energy transfer from
the dark energy to dark matter. Hereafter we assume that the brane
cosmological constant is zero (if it does not vanish, one can absorb
it in the stress-energy tensor of fluid on the brane).

\section{First law of thermodynamics in Vicous braneworld\label{FIRST}}
In this section we are going to examine the first law of
thermodynamics on the brane. In particular, we show that for a
closed universe filled with viscous dark energy and dark matter
the Friedmann equation can be written directly in the form of the
first law of thermodynamics at apparent horizon on the brane.
Using Eq. (\ref{rela}) the Friedmann equation (\ref{RSFri}) can be
written as
\begin{equation}\label{RSAd1}
 \sqrt{H^2+\frac{k}{a^2}+\frac{1}{\ell^2}}= \frac{4\pi
 G_{5}}{3}(\rho_m+\rho_D).
 \end{equation}
In terms of the apparent horizon radius we have
\begin{equation}\label{RSAd2}
 \rho_m+\rho_D = \frac{3}{4\pi
 G_{5}}\sqrt{\frac{1}{{\tilde{r}_A}^2}+\frac{1}{\ell^2}}.
\end{equation}
Taking differential form of equation (\ref{RSAd1}) and using Eqs.
(\ref{consm}) and (\ref{consq}), we can get the differential form
of the Friedmann equation
\begin{equation}\label{RSAd3}
 H\left[\rho_D(1+u+w_D)-3H\xi\right]dt=\frac{\ell}{4\pi
G_5}\frac{d\tilde{r}_A}{\tilde{r}_A^2\sqrt{{\tilde{r}_A}^2+\ell^2}}.
\end{equation}
where $u =\rho_m/\rho_D$ is the ratio of energy densities.
Multiplying both sides of the equation (\ref{RSAd3}) by a factor
$4\pi\tilde{r}_{A}^{3}\left(1-\frac{\dot {\tilde r}_A}{2H\tilde
r_A}\right)$, and using the expression (\ref{surgrav}) for the
surface gravity, after some simplification one can rewrite this
equation in the form
\begin{eqnarray}
\label{RS4} -\frac{\kappa}{2\pi}\frac{2\pi\ell}{G_5}\frac{\tilde
{r}_{A}^{2}d\tilde
{r}_{A}}{\sqrt{{\tilde{r}_A}^2+\ell^2}}&=&4\pi\tilde
 {r}_{A}^{3}H\left[\rho_D(1+u+w_D)-3H\xi\right]dt \nonumber \\
 &&-2\pi\tilde
 {r}_{A}^{2}\left[\rho_D(1+u+w_D)-3H\xi\right]d\tilde {r}_{A}.
 \end{eqnarray}
 $E=(\rho_m+\rho_D) V$ is the total energy content of the universe
inside a $3$-sphere of radius $\tilde{r}_{A}$ on the brane, where
$V=\frac{4\pi}{3}\tilde{r}_{A}^{3}$ is the volume enveloped by
3-dimensional sphere with the area of apparent horizon
$A=4\pi\tilde{r}_{A}^{2}$. Taking differential form of the
relation $E=(\rho_m+\rho_D) \frac{4\pi}{3}\tilde{r}_{A}^{3}$ for
the total matter and energy inside the apparent horizon, we get
\begin{equation}
\label{dE1}
 dE=4\pi\tilde
 {r}_{A}^{2}(\rho_m+\rho_D) d\tilde {r}_{A}+\frac{4\pi}{3}\tilde{r}_{A}^{3}(\dot{\rho}_{m}+\dot{\rho}_{D}) dt.
\end{equation}
Using Eqs. (\ref{consm}) and (\ref{consq}), we obtain
\begin{equation}
\label{dE2}
 dE=4\pi\tilde
 {r}_{A}^{2}\rho_D(1+u) d\tilde {r}_{A}-4\pi  \tilde{r}_{A}^{3}H \left[\rho_D(1+u+w_D)-3H\xi\right]dt.
\end{equation}
Substituting this relation into (\ref{RS4}), after some
simplifications one can rewrite this equation in the form
\begin{equation}\label{Fridif4}
dE -WdV =\frac{\kappa}{2\pi}\frac{2\pi\ell}{G_{5}}\frac{\tilde
{r}_{A}^{2}}{\sqrt{{\tilde{r}_A}^2+\ell^2}}d\tilde {r}_{A}.
\end{equation}
where
$$W=\frac{1}{2}\left[\rho_m+\rho_D-\tilde{p}_D\right]=\frac{1}{2}\rho_D\left[1+u-w_{D}+3H\xi\right],$$
is the matter work density \cite{Hay2}. The work density term is
regarded as the work done by the change of the apparent horizon,
which is used to replace the negative pressure if compared with
the standard first law of thermodynamics, $dE = TdS-pdV$. For a
pure de Sitter space, $\rho_m+\rho_D=-\tilde{p}_D$, then our work
term reduces to the standard $-\tilde{p}_D dV$. Expression
(\ref{Fridif4}) is nothing, but the first law of thermodynamics at
the apparent horizon on the brane, namely $dE=T_hdS_h+WdV$. We can
define the entropy associated with the apparent horizon on the
brane as
\begin{equation}
\label{ent3} S_h=\frac{2\pi\ell}{G_{5}}{\displaystyle\int^{\tilde
r_A}_0\frac{\tilde{r}_A^{2}
}{\sqrt{\tilde{r}_A^2+\ell^2}}d\tilde{r}_A}.
\end{equation}
After the integration we have
\begin{equation} \label{entRSAdS2}
S_h=\frac{2\pi{\tilde{r}_A}^{3}}{3 G_{5}}
 \times
{}_2F_1\left(\frac{3}{2},\frac{1}{2},\frac{5}{2},
-\frac{{\tilde{r}_A}^2}{\ell^2}\right),
\end{equation}
where ${}_2F_1(a,b,c,z)$ is the hypergeometric function. It is
worth noticing when $\tilde{r}_A \ll\ell$, which physically means
that the size of the extra dimension is very large if compared
with the apparent horizon radius, one recovers the $5$-dimensional
area formula for the entropy on the brane
\cite{Shey1,Shey2,Shey3,Shey4}. This is due to the fact that
because of the absence of the negative cosmological constant in
the bulk, no localization of gravity happens on the brane. As a
result, the gravity on the brane is still $5$-dimensional. In this
way we show that for a non-flat universe filled with viscous dark
energy and dark matter the Friedmann equation can be written in
the form of the first law of thermodynamics at apparent horizon in
RSII braneworld.
\section{GSL and interacting viscous dark energy}\label{AdSRS}
Our aim here is to investigate the validity of the generalized
second law of thermodynamics in  a region enclosed by the apparent
horizon on the brane. Taking the derivative of Eq. (\ref{RSAd2})
with respect to the cosmic time  and using Eqs. (\ref{consm}) and
(\ref{consq}), one gets
\begin{equation}
\label{RSAd3} \label{dotr2} \dot{\tilde{r}}_A=\frac{4\pi
}{\ell}G_{5} H{\tilde{r}_A}^2
\left[\rho_D(1+u+w_D)-3H\xi\right]\sqrt{{\tilde{r}_A}^2+\ell^2}.
\end{equation}
Next we turn to calculate $T_{h} \dot{S_{h}}$:
\begin{eqnarray}\label{TSh2}
T_{h} \dot{S_{h}} &=&\frac{1}{2\pi \tilde r_A}\left(1-\frac{\dot
{\tilde r}_A}{2H\tilde r_A}\right)\frac{d}{dt}
\left[\frac{2\pi{\tilde{r}_A}^{3}}{3 G_{5}}
 \times
{}_2F_1\left(\frac{3}{2},\frac{1}{2},\frac{5}{2},
-\frac{{\tilde{r}_A}^2}{\ell^2}\right)\right] \nonumber\\
&=&\frac{1}{2\pi \tilde r_A}\left(1-\frac{\dot {\tilde
r}_A}{2H\tilde
r_A}\right)\frac{2\pi\ell}{G_{5}}\frac{{\tilde{r}_A}^{2}\dot
{\tilde r}_A}{\sqrt{\tilde{r}_A^2+\ell^2}}.
\end{eqnarray}
Using Eq. (\ref{dotr2}), after some simplification we obtain
\begin{equation}\label{TSh3}
T_{h} \dot{S_{h}} =4\pi H\left[\rho_D(1+u+w_D)-3H\xi\right]{\tilde
r_A}^{3}\left(1-\frac{\dot {\tilde r}_A}{2H\tilde r_A}\right).
\end{equation}
As we argued above the term $\left(1-\frac{\dot {\tilde
r}_A}{2H\tilde r_A}\right)$ is positive to ensure $T_{h}>0$,
however, in an accelerating universe the equation of state
parameter of dark energy may satisfy the condition
$w_D<-1-u+3H\xi/\rho_D$. This implies that the second law of
thermodynamics, $\dot{S_{h}}\geq0$, does not hold. However, as we
will see below the generalized second law of thermodynamics,
$\dot{S_{h}}+\dot{S_{m}}+\dot{S_{D}}\geq0$, is still fulfilled
throughout the history of the universe. The entropy of the viscous
dark energy plus dark matter inside the apparent horizon,
$S=S_{m}+S_{D}$, can be related to the total energy
$E=(\rho_m+\rho_D) V$ and pressure $\tilde{p}_D$ in the horizon by
the Gibbs equation \cite{Pavon2}
\begin{equation}\label{Gib1}
T dS=d[(\rho_m+\rho_D) V]+\tilde{p}_DdV=V(
d\rho_m+d\rho_D)+\left[\rho_D(1+u+w_D)-3H\xi\right]dV,
\end{equation}
where $T=T_{m}=T_D$ and $S=S_{m}+S_D$ are, respectively, the
temperature and the total entropy of the energy and matter content
inside the horizon, and $V=\frac{4\pi}{3}\tilde{r}_{A}^{3}$ is the
volume enveloped by the apparent horizon. Here we assumed that the
temperature of both dark components are equal, due to their mutual
interaction. We also limit ourselves to the assumption that the
thermal system bounded by the apparent horizon remains in
equilibrium so that the temperature of the system must be uniform
and the same as the temperature of its boundary. This requires
that the temperature $T$ of the viscous dark energy inside the
apparent horizon should be in equilibrium with the temperature
$T_h$ associated with the apparent horizon, so we have $T=T_h$.
This expression holds in the local equilibrium hypothesis. If the
temperature of the fluid differs much from that of the horizon,
there will be spontaneous heat flow between the horizon and the
fluid and the local equilibrium hypothesis will no longer hold.
This is also at variance with the FRW geometry. In general, when
we consider the thermal equilibrium state of the universe, the
temperature of the universe is associated with the apparent
horizon. Therefore from the Gibbs equation (\ref{Gib1}) we obtain
\begin{equation}\label{TSm1}
T_{h} (\dot{S_{m}}+\dot{S_{D}}) =4\pi {\tilde{r}_{A}^2}
\left[\rho_D(1+u+w_D)-3H\xi\right]\dot{\tilde{r}}_{A}-4\pi H
{\tilde{r}_{A}^3} \left[\rho_D(1+u+w_D)-3H\xi\right].
\end{equation}
To check the generalized second law of thermodynamics, we have to
examine the evolution of the total entropy $S_h + S_m+S_D$. Adding
equations (\ref{TSh3}) and (\ref{TSm1}),  we get
\begin{equation}\label{GSL1}
T_{h}( \dot{S}_{h}+\dot{S}_{m}+\dot{S}_D)=2\pi {\tilde{r}_{A}^2}
\left[\rho_D(1+u+w_D)-3H\xi\right]\dot{\tilde{r}}_A=\frac{A}{2}\left[\rho_D(1+u+w_D)-3H\xi\right]
\dot {\tilde r}_A.
\end{equation}
where $A=4\pi\tilde{r}_{A}^{2}$ is the area of the apparent
horizon on the brane. Substituting $\dot {\tilde r}_A$ from Eq.
(\ref{dotr2}) into (\ref{GSL1}) we reach
\begin{equation}\label{GSL2}
T_{h}( \dot{S}_{h}+\dot{S}_{m}+\dot{S}_D)=\frac{2\pi}{\ell} G_{5}
A {\tilde r_A}^{2}\sqrt{\tilde{r}_A^2+\ell^2}\ H
\left[\rho_D(1+u+w_D)-3H\xi\right]^2.
\end{equation}
The right hand side of the above equation cannot be negative
throughout the history of the universe, which means that $
\dot{S_{h}}+\dot{S_{m}}+\dot{S}_D\geq0$ always holds. This
indicates that the generalized second law of thermodynamics is
fulfilled in the RS II braneworld embedded in the AdS bulk.

\section{GSL and with variable $5$D Newton's
constant}\label{Gvar} In this section we would like to perform the
above analysis with considering the time variable $5$D Newton's
constant $G_5$. There is some evidence of a variable $G_5$ through
numerous astrophysical observations \cite{Ko}. Models with
variable Newton's constant can fix some of the hardest problems in
cosmology like the age problem, cosmic coincidence problem and
finding the value of the Hubble parameter \cite{Gold}.

Taking the derivative of Eq. (\ref{RSAd2}) with respect to the
cosmic time  and using Eqs. (\ref{consm}) and (\ref{consq}), one
gets
\begin{equation}
\label{dotr3v} \dot{\tilde{r}}_A=\left(4\pi G_{5} H
\left[\rho_D(1+u+w_D)-3H\xi\right]-P\dot{G_{5}}\right)\frac{{\tilde{r}_A}^2}{\ell}\sqrt{{\tilde{r}_A}^2+\ell^2}.
\end{equation}
where we have defined
\begin{equation}\label{P}
P=\frac{\sqrt{{\tilde{r}_A}^2+\ell^2}}{{\tilde r}_A G_{5} \ell }.
\end{equation}
Next we calculate $T_{h} \dot{S_{h}}$:
\begin{eqnarray}\label{TSh3v}
T_{h} \dot{S_{h}} &=&\frac{1}{2\pi \tilde r_A}\left(1-\frac{\dot
{\tilde r}_A}{2H\tilde
r_A}\right)\left[\frac{2\pi\ell}{G_{5}}\frac{{\tilde{r}_A}^{2}\dot
{\tilde
r}_A}{\sqrt{\tilde{r}_A^2+\ell^2}}-\frac{\dot{G_{5}}}{G_{5}}S_h\right].
\end{eqnarray}
Next we examine the evolution of the total entropy $S_h +
S_m+S_D$. Adding equations (\ref{TSh3v}) and (\ref{TSm1}),  and
using Eq. (\ref{dotr3v})  we reach
\begin{eqnarray}\label{GSL2v}
T_{h}( \dot{S}_{h}+\dot{S}_{m}+\dot{S}_D)&=&2\pi {\tilde{r}_{A}^2}
\left[\rho_D(1+u+w_D)-3H\xi\right]\dot{\tilde{r}}_A\nonumber
\\&&-\frac{\dot{G_{5}}}{G_{5}}\left[P\tilde{r}_A^3+\frac{S_h}{2\pi
\tilde{r}_{A}}\right]\left(1-\frac{\dot {\tilde r}_A}{2H\tilde
r_A}\right).
\end{eqnarray}
Substituting $\dot {\tilde r}_A$ from Eq. (\ref{dotr3v}) into
(\ref{GSL2v}) we reach
\begin{eqnarray}\label{GSL3v}
T_{h}( \dot{S}_{h}+\dot{S}_{m}+\dot{S}_D)&=&\frac{2\pi}{\ell}
G_{5} A {\tilde r_A}^{2}\sqrt{\tilde{r}_A^2+\ell^2} H
\left[\rho_D(1+u+w_D)-3H\xi\right]^2 \nonumber \\
&&-\frac{A}{2}P
\dot{G_{5}}\left[\rho_D(1+u+w_D)-3H\xi\right]\frac{\tilde
r_{A}^{2}}{\ell} \sqrt{\tilde{r}_A^2+\ell^2}\nonumber \\
&&-\frac{\dot{G_{5}}}{G_{5}}\left[P\tilde{r}_A^3+\frac{S_h}{2\pi
\tilde{r}_{A}}\right]\left(1-\frac{\dot {\tilde r}_A}{2H\tilde
r_A}\right).
\end{eqnarray}
For $\dot{G_{5}}=0$, we obtain the result of the previous section.
In this case the validity of GSL depend to the sign of
$\dot{G_{5}}$, if $\dot{G_{5}}<0$, and $\rho_D(1+u+w_D)>3H\xi$, then
$T_{h}( \dot{S}_{h}+\dot{S}_{m}+\dot{S}_D)>0$.
\section{ Summary and discussions}\label{sum}
In the present, paper we have showed that the Friedmann equations on
a RSII braneworld filled with interacting viscous dark energy and
dark matter can be written directly in the form of the first law of
thermodynamics at apparent horizon. Then We examined the validity of
the generalized second law of thermodynamics, we studied the time
evolution of the total entropy, including the entropy associated
with the apparent horizon and the entropy of the viscous dark energy
inside the apparent horizon. Our study have shown that the
generalized second law of thermodynamics is always protected in a
RSII braneworld filled with interacting viscous dark energy and dark
matter in a region enclosed by the apparent horizon. Then we
extended our study to the case time variable 5-dynamical Newton's
constant $G_5$. According to the our calculations the generalized
second law of thermodynamics is valid if $\dot{G_{5}}<0$, and
$\rho_D(1+u+w_D)>3H\xi$. These results hold regardless of the
specific form of the dark energy.
\acknowledgments{This work has been supported by Research Institute
for Astronomy and Astrophysics of Maragha.}


\begin{thebibliography}{99}
\bibitem{1}A.~G.~Riess {\it et al.}  [Supernova Search Team Collaboration],
  Astron.\ J.\  {\bf 116}, 1009 (1998);\\
S. Perlmutter {\it et al.} [Supernova Cosmology Project
Collaboration], Astrophys. J. {\bf 517}, 565 (1999).

\bibitem{111}D. N. Spergel, Astrophys. J. Suppl. 148 (2003) 175;\\
C. L. Bennett, et al.,  Astrophys. J. Suppl. 148 (2003) 1; \\ U.
Seljak, A. Slosar, P. McDonald, JCAP 0610 (2006) 014;\\ D. N.
Spergel, et al., Astrophys. J. Suppl. 170 (2007) 377. .
\bibitem{2}E. J. Copeland, M. Sami and Shinji Tsujikawa, Int. J. Mod. Phys. D 15 (2006) 1753, [
arXiv:hep-th/0603057]; N. Straumann, [arXiv:astro-ph/0009386]; Y.
Fujii, Phys. Rev. D 62 (2000) 064004; L. P. Chimento, A. S. Jakubi
and D. Pavon, [arXiv:astro-ph/0010079];  J. Kujat, R. J. Scherrer
and A. A. Sen, Phys. Rev. D 74 (2006) 083501; Y. -F. Cai, E. N.
Saridakis, M. R. Setare, J. -Q. Xia, arXiv:0909.2776 [hep-th].

\bibitem{3}P. J. E. Peebles and B. Ratra, Rev. Mod. Phys. 75 (2003)
559.
\bibitem{4}F. Perrotta, C. Baccigalupi and S. Matarrese, Phys. Rev. D. B61 (2000) 023507; B.
Boisseau, G. Esposito-Farese, D. Polarski and A. A. Starobinsky,
Phys. Rev. Lett. 85 (2000) 2236; M. R. Setare, Phys. Lett. B
{\bf644}, 99, (2007).

\bibitem{5}S. Capozziello, V. F. Cardone, S. Carloni and A. Troisi, Int. J. Mod. Phys. D 12 (2003)
1969-1982, [arXiv:astro-ph/0307018]; S. Carroll et al. Phys. Rev. D
70 (2004) 043528; S. Nojiri and S. D. Odintsov, AIP Conf. Proc. 1115
(2009) 212-217, [arXiv:0810.1557]; T. P. Sotiriou and V. Faraoni, [
arXiv:/0805.1726]; A. A. Starobinsky, JETP. Lett. 86(2007) 157,
[arXiv:/0706.2041];  M. R. Setare, Int. J. Mod. Phys. D 17 (2008)
2219, [arXiv:0901.3252]; M. R. Setare, arXiv:0908.0196 [gr-qc].

\bibitem{6}G. Dvali, G. Gabadadze andM. Porrati, Phys. Lett. B 485 (2000) 208, [hep-th/0005016];P. Moyassari and
M. R. Setare, Phys. Lett. B 674 (2009) 237, [ arXiv:0806.2418]; M.
R. Setare, E. N. Saridakis, JCAP 0903: 002, (2009); M. R. Setare,
Int. J. Mod. Phys. D{\bf18}, 419, (2009); M. R. Setare,  J. Sadeghi,
A. R. Amani, Phys. Lett. B{\bf 660}, 299, (2008).


\bibitem{7}D. J. Gross and J. H. Sloan, Nucl. Phys. B 291 (1987) 41; C. Charmousis and J. F.
Dufaux, Class. Quant. Grav. 19 (2002) 4671, [arXiv:hepth/ 0202107];
S. C. Davis, Phys. Rev. D 67 (2003) 024030, [arXiv:hep-th/0208205];
M.R. Setare,  J. Sadeghi,A. R. Amani, Int. J. Mod. Phys. D{\bf 18},
1291, (2009).
\bibitem{Pad}T. Padmanabhan and S. M. Chitre, Phys. Lett. A 120, 433 (1987).
\bibitem{Bre0}I. Brevik and L. T. Heen, Astrophys. Space Sci. 219, 99 (1994);\\
 Brevik and A. Hallanger, Phys. Rev. D 69, 024009 (2004).
\bibitem{Bre1} I. Brevik and O. Gorbunova, Gen. Relativ. Gravit. 37, 2039
(2005).

\bibitem{Bre2} I. Brevik, O. Gorbunova and Y. A. Shaido, Int. J. Mod.
Phys. D 14, 1899 (2005);\\
I. Brevik and O. Gorbunova, Eur. Phys. J. C 56, 425 (2008);\\ I.
Brevik, Eur. Phys. J. C 56, 579 (2008).
\bibitem{G4com}
S.~D'Innocenti, G.~Fiorentini, G.~G.~Raffelt, B.~Ricci and A.~Weiss,
Astron. Astrophys.  {\bf 312}, 345 (1996); K.~Umezu, K.~Ichiki and
M.~Yahiro, Phys. Rev. D {\bf 72}, 044010 (2005);
 S.~Nesseris and L.~Perivolaropoulos, Phys. Rev. D {\bf 73}, 103511
(2006); J.~P.~W.~Verbiest {\it et al.} [arXiv:astro-ph/0801.2589].

\bibitem{kogan}
  G.~S.~Bisnovatyi-Kogan,
  Int.\ J.\ Mod.\ Phys.\  D {\bf 15}, 1047 (2006).


\bibitem{Damour}
Damour T.,{\it  et al}, Phys. Rev. Lett. {\bf 61}, 1151 (1988).

\bibitem{guenther} D.B. Guenther, Phys. Lett. B {\bf 498}, 871 (1998).

\bibitem{Gaztanaga}
  E.~Gaztanaga, E.~Garcia-Berro, J.~Isern, E.~Bravo and I.~Dominguez,
  Phys.\ Rev.\  D {\bf 65}, 023506 (2002).


\bibitem{Biesiada} Biesiada M. and Malec B.,  Mon. Not. R. Astron. Soc. {\bf 350}, 644
(2004).



\bibitem{ray1}
  S.~Ray and U.~Mukhopadhyay,
  Int.\ J.\ Mod.\ Phys.\  D {\bf 16}, 1791 (2007).


\bibitem{gol}I. Goldman, Phys. Lett. B {\bf281}, 219 (1992).

\bibitem{jamil}
  M.~Jamil, F.~Rahaman and M.~Kalam,
  Eur.\ Phys.\ J.\  C {\bf 60}, 149 (2009).



\bibitem{ber}O. Bertolami et al, Phys. Lett. B {\bf311}, 27 (1993).

\bibitem{Dirac:1938mt}
  P.~A.~M.~Dirac,
  Proc.\ Roy.\ Soc.\ Lond.\  A {\bf 165} (1938) 199.

\bibitem{kal} T. Kaluza, Sitz. d. Preuss. Akad. d. Wiss. Physik-Mat. Klasse (1921),
966.

\bibitem{akk}P. G. O. Freund, Nuc. Phys. B. {\bf 209}, 146 (1982); K. Maeda, Class. Quant. Grav. {\bf 3}, 233 (1986);
 E. W. Kolb, M. J. Perry and T. P. Walker, Phys. Rev. D {\bf
33},869 (1986); P. Lor´e-Aguilar, E. Garc´i-Berro, J. Isern, and Yu.
A. Kubyshin, Class. Quant. Grav. {\bf 20}, 3885 (2003).

\bibitem{bd} C. H. Brans and R. H. Dicke, Phys. Rev. {\bf 124} (1961) 925.

\bibitem{gen} P. G. Bergmann, Int. J. Theor. Phys. {\bf 1} (1968), 25; R. V. Wagoner, Phys. Rev. D {\bf 1} (1970),
3209; K. Nordtvedt, Astrophys. J. {\bf 161} (1970), 1059.

\bibitem{bek} J. D. Bekenstein, Found. Phys. {\bf 16}, 409 (1986).

\bibitem{19}
 I. L.
Shapiro and J. Sola, JHEP 0202 (2002) 006; A. Babi\'c, B. Guberina,
R. Horvat, and H. \v{S}tefan\v{c}i\'c, Phys. Rev. D {\bf 65}, 085002
(2002);  I. L. Shapiro, J. Sola, C. Espana-Bonet, and P.
Ruiz-Lapuente, Phys. Lett. B {\bf 574}, 149 (2003); B. Guberina, R.
Horvat, and H. \v{S}tefan\v{c}i\'c, Phys. Rev. D {\bf 67}, 083001
(2003);
 C. Espana-Bonet, P. Ruiz-Lapuente, I. L. Shapiro and J.
Sola, JCAP {\bf 0402}, 006 (2004).


\bibitem{21} M. Reuter, Phys. Rev. D {\bf 57} (1998) 971;
A. Bonnano and M. Reuter, Phys. Rev. D {\bf65} 043508 (2002).

\bibitem{71}
  R.~Horvat,
  Phys.\ Rev.\  D {\bf 70}, 087301 (2004);

\bibitem{Guberina}
B.~Guberina, R.~Horvat and H.~Nikolic, Phys. Rev. D {\bf 72}, 125011
(2005).
\bibitem{wmap}C. L. Bennett
et al., Astrophys. J. Suppl. 148, 1 (2003); D. N. Spergel,
Astrophys. J. Suppl. 148, 175, (2003).
\bibitem{Hay2} S.A. Hayward, S. Mukohyana, and M. C. Ashworth, Phys.
Lett.  A {\bf 256}, 347 (1999);\\ S. A. Hayward, Class. Quantum
Grav. {\bf 15}, 3147 (1998)

\bibitem{Bak} D. Bak and S. J. Rey, Class. Quantum Grav. {\bf17}, L83 (2000).
\bibitem{cai3} R.~G.~Cai and L.~M.~Cao, Y. P. Hu, arXiv:0809.1554.

\bibitem{Bin} P. Binetruy, C. Deffayet, and D. Langlois, Nucl.
Phys. B {\bf565} (2000) 269.

\bibitem{Zim} W. Zimdahl and D. Pav´on, Phys. Rev. 61, 108301 (2000).

\bibitem{Shey1} A. Sheykhi, B. Wang and R. G. Cai, Nucl. Phys. B {\bf
779} (2007)1.
  \bibitem{Shey2} A. Sheykhi, B. Wang and R. G. Cai, Phys. Rev. D {\bf
76} (2007) 023515.

\bibitem{Shey3} A. Sheykhi, B. Wang, Phys. Lett. B 678 (2009) 434.

  \bibitem{Shey4} A. Sheykhi, B. Wang, arXiv:0811.4477.




\bibitem{Ko} G.S.B. Kogan, gr-qc/0511072;\\ D.B. Guenther, Phys. Lett. B
498, 871 (1998);\\ S. Ray, U. Mukhopadhyay, astro-ph/0510549.


\bibitem{Gold} I. Goldman, Phys. Lett. B 281, 219 (1992);\\ O. Bertolami et
al., Phys. Lett. B 311, 27 (1993); \\ A.I. Arbab,
hep-th/0711.1465v1




\bibitem{Pavon2} G. Izquierdo and D. Pavon, Phys.Lett. B633 (2006)
420.
\end{thebibliography}
\end{document}